# Origin of superconductivity in hole doped SrBiO$_3$ bismuth oxide perovskite from parameter-free *first-principles* simulations

Julien Varignon[1]

[1] Laboratoire CRISMAT, CNRS UMR 6508, ENSICAEN, Normandie Université, 6 boulevard Maréchal Juin, F-14050 Caen Cedex 4, France

**Abstract**

The recent discovery nickel oxides superconductors have highlighted the importance of *first-principles* simulations for understanding the formation of the bound electrons at the core of superconductivity. Nevertheless, superconductivity in oxides is often ascribed to strong electronic correlation effects that Density Functional Theory (DFT) cannot properly take into account, thereby disqualifying this technique. Being isostructural to nickel oxides, Sr$_{1-x}$K$_x$BiO$_3$ superconductors form an ideal testbed for unveiling (i) the lowest theory level needed to model complex superconductors and (ii) the underlying pairing mechanism yielding superconductivity. Here I show that parameter-free DFT simulations capture all the experimental features and related quantities of Sr$_{1-x}$K$_x$BiO$_3$ superconductors, encompassing the prediction of an insulating to metal phase transition upon increasing the K doping content and of an electron-phonon coupling constant of 1.22 in sharp agreement with the experimental value of 1.3 +/- 0.2. Proximity of a disproportionated phase is further demonstrated to be a prerequisite for superconductivity in bismuthates. This work highlights that appropriately executed DFT is a sufficient platform for studying superconductivity in complex oxides.

## INTRODUCTION

Superconductivity is a peculiar state of materials characterized by zero resistance to direct current and expulsion of magnetic flux. It is explained by the formation of bound electrons called Cooper pairs[1]. Up to date, the microscopic mechanism behind the Cooper pair formation is yet to be clarified and unified between all known superconductors. In simple elements, it is usually explained by exchange of phonons. In high temperature oxides superconductors, the proximity of a magnetic phase transition and/or a charge ordered state is proposed to explain the formation of bound electrons. The discovery of superconductivity in nickel based oxides $R_{1-x}Sr_xNiO_2$ in 2019[2] arouses newest interest from solid state scientists as it offers a new testbed for theories of superconductivity in complex oxides. It also highlights the importance of electronic structure calculations for understanding phenomena associated with superconductivity. In that regard, there is an established consensus that high critical temperature ($T_c$) reached in the oxide superconductors might be favored by strong correlation effects that have to be accounted in electronic structure simulations[3–9].

Aiming at understanding the role of electronic correlations and the mechanism behind Cooper pairs formation, $Sr_{1-x}K_xBiO_3$ and $Ba_{1-x}K_xBiO_3$ sit as ideal compounds for testing our *first-principles* simulations techniques since these materials (i) host almost all complexity exhibited by oxides and (ii) are the only oxide superconductors adopting the simple $ABO_3$ perovskite structure[10–12]. In the bulk, $SrBiO_3$ and $BaBiO_3$ are insulating with a band gap $E_g$ estimated between 0.2 and 0.8 eV in $BaBiO_3$ [13–15]. They both adopt a monoclinic *P2$_1$/n* symmetry at low temperature[10,16], characterized by the usual $a^0a^0c^+$ ($\emptyset_z^+$) and $a^-a^-c^0$ ($\emptyset_{xy}^-$) octahedra rotations in Glazer's notations[17] (Figs. 1.a and b) induced by the A-to-B cation size mismatch – quantified by a Goldschmidt[18] tolerance factor t = 0.88 in $SrBiO_3$. While the two $O_6$ group rotations produce the usual orthorhombic *Pbnm* symmetry exhibited by most $ABO_3$ perovskites, the *P2$_1$/n* phase is reached by the appearance of a breathing distortion $B_{oc}$ (Fig 1.c) – also called bond disproportionation[19] – producing a dimerization of the material along the [111] cubic direction. This lattice instability, appearing at the R point of the primitive *Pm-3m*, undistorted, cubic cell Brillouin zone (*i.e.* (1/2,1/2,1/2)), results in a rock-salt pattern of large and compressed $O_6$ octahedra, splitting the Bi cations into $Bi_L$ and $Bi_S$, respectively (Fig. 1.c)[20,21]. The breathing mode is associated with disproportionation of the unstable 4+ formal oxidation state (FOS) of $Bi^{4+}$-$6s^1$ cations to more stable 3+ ($6s^2$) and 5+ ($6s^0$) FOS in the *P2$_1$/n* insulating phase[14,19,20,22,23]. However, bismuthates fall within the negative charge transfer insulator regime[24,25] and the Bi *6s* band is localized well below the Fermi level and the O *2p* band[22,23,26]. While the disproportionation should result in the localization of spin paired electrons and holes on the *s* states of $Bi_L$ and $Bi_S$ cations, respectively, O anions supply electrons to the depleted $Bi_S^{5+}$ cations resulting in the localization of two holes on the surrounding O atoms and yielding a $Bi_L$–*6s²* and $Bi_S$–*6s²L²* electronic configuration where the notation *L* stands for a ligand hole[19,22,23,26,27].

Upon hole doping by substituting the divalent Sr or Ba cations by the monovalent K ion, $Sr_{1-x}K_xBiO_3$ and $Ba_{1-x}K_xBiO_3$ show superconductivity for doping contents ranging from x=0.45-0.6 and 0.3-0.45, respectively, with a critical temperature $T_c$ measured between 5 to 30 K, depending on the doping content as well as on the O stoichiometry of the samples[10–12]. Superconductivity is explained by an electron-phonon coupling (EPC) with a constant $\lambda = 1.3 \pm 0.2$ [4]. It is proposed on the basis of *first-principles* simulations that $\lambda$ is enhanced by strong electronic correlations[3,4] but also by octahedral rotations[28]. At moderate hole doping content, these compounds exhibit a semiconducting state, explained by the appearance of localized states in the gap and trapped hole in the lattice (*i.e.* a hole bipolaronic state)[27,29].

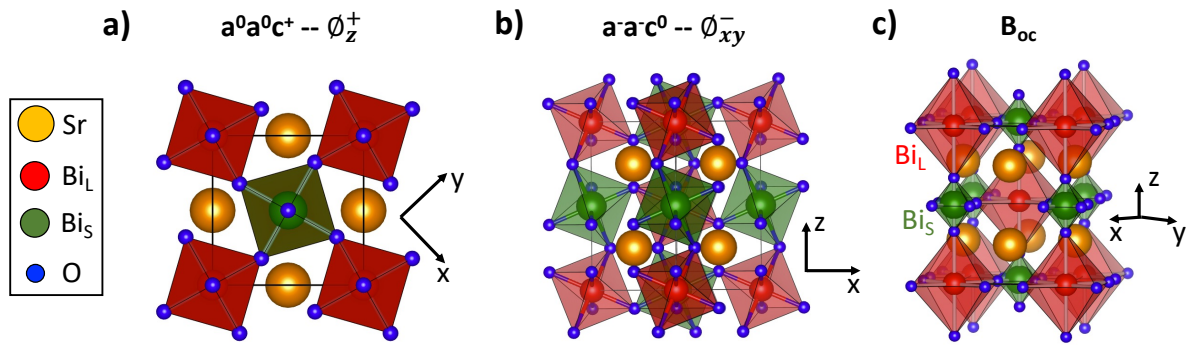

*Figure 1 : Structural distortions displayed by SrBiO₃ and BaBiO₃ in the P2₁/n ground state phase. a) In phase rotation a⁰a⁰c⁺ ($\phi_z^+$) around the z axis. b) Anti phase rotation a⁻a⁻c⁰ ($\phi_{xy}^-$) around the x and y axes. c) Breathing distortion $B_{oc}$ splitting Bi sites into two different types.*

Several theoretical studies have tried to address the physics of the bismuthates and their related superconducting phase[3,19,30–35,20–23,26–29]. Most of these studies point that Density Functional Theory (DFT) is unable to account for the insulating character or appearance of the breathing mode $B_{oc}$ in $BaBiO_3$ within the usual Local Density Approximation (LDA) and Generalized Gradient Approximation (GGA) exchange-correlation functionals[3,20,21,36] thereby hindering its doping effect study. This is also true for the extraction of superconducting quantities such as electron-phonon matrix elements and coupling strength that are underestimated by LDA or GGA with respect to experiments ($\lambda$ =0.34[3,21] or 0.48[28] with LDA and GGA, respectively, instead of 1.3 with Angular Resolved Photo Emission Spectroscopy experiments[4]). Both the band gap and superconducting fundamental quantities can be improved by more sophisticated, but prohibitive, DFT hybrid functional and/or Green Functions and screened Coulomb interaction (GW) calculations[3,27,29]. Nevertheless, including all degrees of freedom (*e. g.* structural such as symmetry lowering events and cation disorder appearing in alloys, electronic instabilities lifting degeneracies , local spin formation and long range spin orders…) of complex oxides for different doping contents are not affordable for these techniques, that are way too demanding in terms of computational resources since one has to deal with very large supercells – hybrid DFT and GW calculations are usually restricted to very small cell size preventing symmetry lowering events. Thus, affordable but still predictable DFT calculations are required for understanding trends in doping of complex oxides and superconductors.

Recently, the band gap opening of $ABO_3$ oxides perovskites with a *3d* element has been addressed by DFT simulations and revealed to originate from four simple modalities rather than the usual explanation based on strong dynamical correlation effects codified by the Hubbard model[37]: (i) the octahedral crystal field splitting the *d* states and Hund's rule; (ii) symmetry lowering events such as octahedral rotation further lifting orbital degeneracies; intrinsic electronic instabilities yielding (iii) a Jahn-Teller effect and removing orbital degeneracies or (iv) disproportionation effects of unstable formal oxidation state (FOS) to more stable FOS and resulting in a double local environment for the B cations. These results have been ratified by using an appropriate DFT exchange-correlation functionals properly amending self-interactions errors inherent to the implementation of DFT and by supplying enough flexibility to the simulation (*e.g.* local motif, symmetry lowering events, spin polarization…)[19,25,37–39]. Furthermore, these conclusions are supported by DFT simulations using the Strongly Constrained and Appropriately Normalized (SCAN)[40] exchange-correlation functional, but without any empirical parameter U such as in DFT+U, that properly account for the bulk perovskite oxide properties[25,38] and the trends in doping effects of oxide insulators such as rare-earth nickelates[41] $RNiO_3$ or copper oxides[42] such as $La_2CuO_4$. One may thus question "*what is the lowest DFT exchange-correlation functional needed for capturing trends in doping effects in complex oxides superconductors*". Furthermore, the bismuthates share several similarities with the rare-earth nickelates $RNiO_3$ in terms of structural and electronic properties (disproportionation effects, similar critical temperature in the 2D superconducting form $RNiO_2$), and thus one may wonder if "*one can extract trends and gain insights on the physics and pairing mechanisms of doped bismuthates that one can potentially transfer to the superconducting nickelates and other oxide superconductors*".

Here I show that the SCAN functional is sufficient to capture the trends in insulating to metal character of $SrBiO_3$ upon hole doping and to reveal the mechanism and prerequisites behind the appearance of superconductivity. By mapping the *first-principles* DFT results on a Landau model involving the relevant lattice distortions, the insulating phase is shown to be reached by disproportionation effects associated with an intrinsic instability of $Bi^{4+}$ cations to disproportionation to $Bi^{3+}/Bi^{5+}$ cations in the bulk ground state. This accompanied by a breathing mode distortion $B_{oc}$ whose amplitude is further enhanced by the octahedra rotations. At weak doping content (x=0.0625 to 0.125), holes are trapped on the lattice and intermediate states are localized in the band gap, ultimately resulting in a semi-conducting behavior. At intermediate doping content (x=0.1875 to 0.375), a metallic phase is reached but the breathing mode is still present in the ground state due to its coupling with octahedral rotations, despite the fact that the structural distortion alone is not willing to spontaneously pop up in the material. The presence of this mode induces small gaps in the bands dispersing around the Fermi level. No gaps are anymore identified in the band structure at x=0.4375, a doping content reminiscent of the superconducting phase reported experimentally (x=0.45-0.6). Around x≥0.4375, the breathing mode is found on the verge of becoming stable in the material due to octahedral rotations and thus its vibration can form spin paired electrons and holes in the material, *i.e.* Cooper pairs. At larger doping content, the breathing mode becomes extremely hard and its vibration becomes unlikely, hindering Cooper pairs formation. These results thus suggests that the proximity of a lattice instability producing spin paired electrons and holes is a prerequisite for superconductivity in the bismuthates, in sharp agreement with the bounded doping content observed experimentally for the superconducting phase. Within the superconducting phase at x=0.4375, an electron-phonon coupling constant $\lambda$ associated with the breathing mode $B_{oc}$ of 1.22 and a $B_{oc}$ frequency of 66 meV are extracted from the simulations, in sharp agreement with the experimental values ($\lambda$ =1.3$\pm$0.2 and $\omega$=62 meV, respectively). This study thus (i) validates the use of SCAN-DFT for studying doping effects in complex oxide superconductors and (ii) calls for inspection of disproportionation effects in superconducting nickelates and other oxide superconductors to see if the identified mechanism in bismuthates is also relevant for these newly identified oxide superconductors.

**RESULTS**

**1. The bulk material**

***DFT ground states properties:*** The bulk *P2$_1$/n* structure experimentally observed at low temperature for SrBiO$_3$ is first relaxed in order to identify the ground state with our DFT functional. Key structural and electronic properties of the bulk material are reported in Table 1. SCAN-DFT predicts that SrBiO$_3$ is an insulator in agreement with experiments, with a band gap amplitude improved with respect to GGA calculations (0.3 eV in Ref.[23] instead of 0.48 in the present study). In terms of structural parameters, the computed lattice parameters are in close agreement with experiments[10], with less than 1% of error on the volume of the unit cell. Regarding the key lattice distortion amplitude exhibited by the compound, the optimized structure in DFT is also in agreement with the experimental structure although the amplitude of the breathing mode $B_{oc}$ is underestimated by ~10% with respect to experiment. This mode produces the rock-salt pattern between the compressed and contracted octahedra (see Figure 1.c). It results in a clear cut of the electronic structure between the two type of Bi cations as inferred by the projected density of states of Figure 2.a. Nevertheless, the Bi-*s* states are well below the O-*p* states, the band gap is mostly formed between occupied and unoccupied O-*p* states (see Figure 2.a) and SrBiO$_3$ falls within the negative charge transfer insulator regime. Using the partial charge density maps of states just above the Fermi level (Fig 2.b), the unoccupied O-*p* bands have a *s*-like orbital character centered on Bi$_s$ site – in agreement with the Wannier functions analysis presented in Ref.[23] –, hinting at the fact that the O anions supply electrons to the depleted Bi$_s$ cation and bear the two holes that are centered at Bi$_s$ sites.

|         | a (Å) | b (Å) | c (Å) | beta (°) | $\phi^-_{xy}$ (Å/f.u.) | $\phi^+_z$ (Å/f.u) | $B_{oc}$ (Å/f.u.) | $E_g$ (eV) |
|---------|-------|-------|-------|----------|------------------------|--------------------|-------------------|------------|
| SCAN-DFT | 5.951 | 6.140 | 8.491 | 90.051 | 0.930 | 0.577 | 0.158 | 0.48 |
| Exp.     | 5.948 | 6.095 | 8.485 | 90.063 | 0.837 | 0.581 | 0.175 | Ins. |

*Table 1 : Key structural lattice and electronic parameters from DFT and comparison with experiments. Experimental results are taken from Ref.[10]. Amplitudes of lattice distortions (in Å/f.u.) are obtained by performing a symmetry mode analysis with respect to a cubic cell whose lattice parameter is fixed so that the cubic cell volume equals the relaxed P2$_1$/n cell volume.*

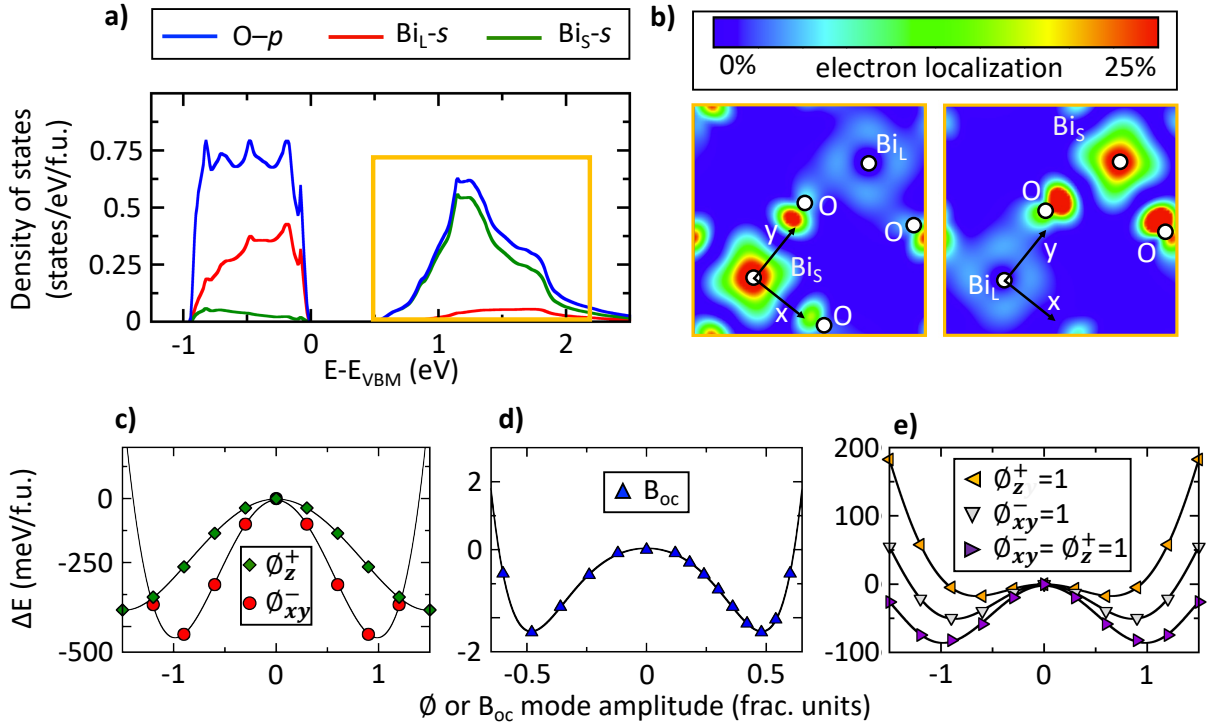

*Figure 2: Ground state properties of SrBiO$_3$. a) Density of states (in states/eV/f.u) projected on Bi$_S$-s (in green), Bi$_L$-s (in red) and O-p (in blue) states. b) Partial charge density map associated with the states at the bottom of the conduction band indicated by the orange area in panel a. c-e) Potential energy surface starting from a perfectly undistorted Pm-3m cubic cell associated with the anti-phase a$^-$a$^-$c$^0$ $\phi^-_{xy}$ and in-phase a$^0$a$^0$c$^+$ $\phi^+_z$ octahedral rotations (d), breathing mode $B_{oc}$ (e) and breathing mode $B_{oc}$ at fixed $\phi^+_z$ and $\phi^-_{xy}$ octahedral rotations amplitude. The lattice parameter of the cubic cell is chosen so that its volume is identical to*

*the ground state volume. An amplitude of 1 corresponds to the amplitude appearing in the DFT relaxed ground state structure.*

**The origin of the insulating phase:** Figure 2.c-e report the potential energy surface associated with the relevant distortion modes identified in the ground state of SrBiO$_3$ when starting from a perfectly undistorted *Pm-3m* cubic cell (see method). As one can see, octahedral rotations show a double well potential whose minimum is located at non-zero amplitudes $Q$ of the distortions. It indicates that these modes are unstable and are willing to spontaneously appear in the ground state due to steric effects. The breathing mode B$_{oc}$ also exhibits a double well potential, albeit with smaller energy gains with respect to octahedral rotations. (Fig 2.d) Thus, the material spontaneously wants to get rid of the unstable 4+ formal oxidation state (FOS) of Bi cations and to adopt the more stable 3+/5+ FOS. This is in agreement with previous DFT work on BaBiO$_3$ performed by Tonhauser and Rabe in Ref.[20]. Nevertheless, the breathing mode would exhibit only 50% of its amplitude appearing in the *P2$_1$/n* ground state if considered alone.

**The importance of octahedral rotations for the stabilization of the breathing mode B$_{oc}$:** Mercy *et al* identified in rare-earth nickelates RNiO$_3$ (R=Lu-Pr, Y)[43] that the breathing mode B$_{oc}$ amplitude $Q_{Boc}$ is directly connected to the octahedra rotation amplitudes $Q_{\phi_{xy}^-}$ and $Q_{\phi_z^+}$, respectively, through the lattice mode couplings allowed in the free energy expansion $F$ displayed in eq.1 :

$$F \propto aQ_{Boc}^2 + bQ_{Boc}^2 Q_{\phi_{xy}^-}^2 + cQ_{Boc}^2 Q_{\phi_z^+}^2 \propto (a + bQ_{\phi_{xy}^-}^2 + cQ_{\phi_z^+}^2)Q_{Boc}^2 = a_{eff} Q_{Boc}^2 \quad \text{(eq.1)}$$

where a, b and c are coefficients. It follows that rotations can renormalize the coefficient in front of $Q_{Boc}^2$ thereby acting on the possibility to stabilize a finite amplitude of B$_{oc}$ ($a_{eff}$<0) or not ($a_{eff}$>0) in the material. Figure 2.e displays the potential energy surface as a function of the B$_{oc}$ mode amplitude $Q_{Boc}$ but at fixed octahedral rotation amplitudes $Q_{\phi_{xy}^-}$ and $Q_{\phi_{xy}^-}$. The stabilization of the breathing mode is enhanced by the O$_6$ groups rotations, increasing the amplitude of the B$_{oc}$ mode appearing in the material and associated energy gain. Obviously, even though the breathing mode B$_{oc}$ alone would not be intrinsically unstable (*a*>0 in eq.1), it can still be forced to appear in the ground state due to its coupling with finite octahedra rotation amplitude yielding *a$_{eff}$*<0 in eq.1. Thus, neglecting octahedral rotations by using smaller, but more convenient unit cells, is not indicated as it can ultimately act on the presence of $B_{oc}$ and related electronic features.

## 2. Polaronic state formation upon hole doping

***Intermediate states in the band gap:*** Following the understanding of the bulk properties, DFT calculations at low hole doping content have been performed in order to check whether or not polaronic states can be trapped on the lattice and yield a semiconducting behavior. To that end, a $Sr_{1-x}K_xBiO_3$ solid solution with x=0.0625 using a 32 f.u. supercell is considered (see method). Then, 2 $Sr^{2+}$ cations are substituted by 2 $K^+$ cations in the 32 f.u. supercell, yielding the release of two holes in the material. After the structural relaxation, a semiconducting state with a band gap of 0.25 eV is identified in the material. By inspecting the projected density of states presented in Figure 3.a, a split-off band localized in the band gap mostly formed by O-*p* and $Bi_L$-*s* states, very similar to the character of the valence band maximum, is revealed. An acceptor state is thus created in the material. The charge density maps associated with this intermediate state presented in Fig. 3.b confirms that the holes are localized on a $Bi_L$ cation that is occupied by 2 spin paired electrons in the pristine material.

***Existence of a trapped hole bipolaronic state:*** This is locally accompanied by a modification of the lattice with the average "$Bi_L$-O bond length" of 2.16 Å on this hole site while $Bi_L$-O bond lengths are roughly 2.30 Å for the other $Bi_L$ sites in the material. It is in fact very similar to a $Bi_S$ cation that indeed shows an average $Bi_S$-O bond length of 2.15 Å. In conclusion, a hole bipolaronic state is trapped in the material and tends to transform $Bi_L$ cations into $Bi_S$ cations, *i.e.* locally annihilating the breathing distortion mode $B_{oc}$. This is in sharp agreement with previous DFT predictions in hole doped $BaBiO_3$ performed with higher DFT functionals[29] that found bipolaronic states at low doping contents.

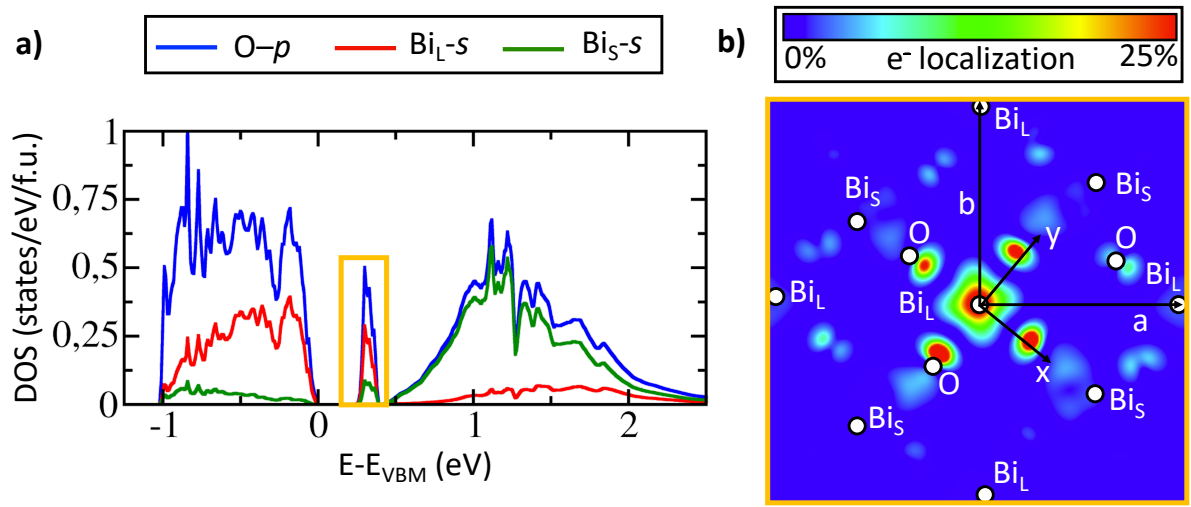

*Figure 3 : Electronic properties of Sr$_{0.9375}$K$_{0.0625}$BiO$_3$.* a) Projected density of states on O-p (in blue), Bi$_L$-s (in red) and Bi$_S$-s (green) states. The zero is set at the valence band maximum (VBM). The orange area indicates the intermediates states localized in the band gap. b) Partial charge density maps associated with the intermediate states localized in the band gap corresponding to the orange area in panel a.

### 3. Trends in insulating-to-metal transition and structural properties upon hole doping

***The insulator to metal transition upon K substitutions*:** The projected density of states on Bi$_L$ and Bi$_S$-*s* states as a function of various K doping content x are reported in Figure 4.a. At moderate doping content (x=0.0625 and x=0.125), intermediate states are formed in the gap and yield a semiconducting behavior. At x=0.1875 up to x=0.375, the material is found metallic but still with a clear asymmetry of electronic structures between Bi$_L$ and Bi$_S$ cations. This hints at the fact that parts of disproportionation may still be present in the material. Finally, at x=0.4375 up to x=0.625, no asymmetries between Bi$_S$ and Bi$_L$ electronic structures are observed suggesting that disproportionation effects have disappeared and the two Bi sites become equivalent in the material.

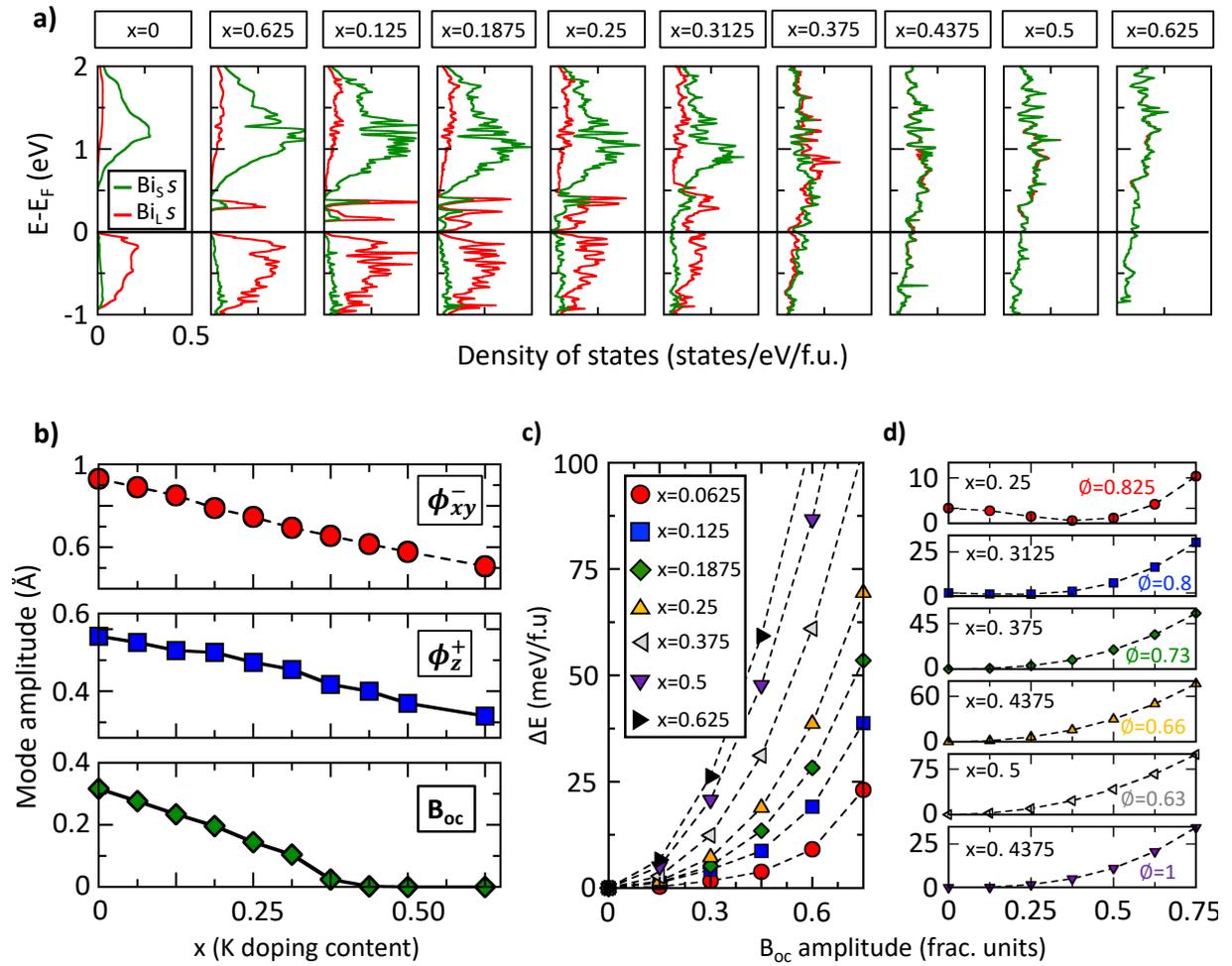

*Figure 4 : Trends in electronic and structural properties upon hole doping SrBiO₃.* a) Projected density of states on $Bi_S$ (in green) and $Bi_L$ (in red) s states as a function of the K doping content x. The Fermi level in semiconducting states (0%, 6.25% and 12.5 %) is set to the valence band maximum. b) Amplitudes of lattice distortions (in Å/f.u) extracted from a symmetry mode analysis for the $a^-a^-c^0$ ($\phi_{xy}^-$) octahedral rotation (upper panel), $a^0a^0c^+$ ($\phi_z^+$) octahedral rotation (middle panel) and breathing mode ($B_{oc}$) (lower panel). c) Potential energy surface (in meV/f.u) associated with the $B_{oc}$ mode amplitude (in fractional units) starting from a perfectly cubic cell for different doping content x. d) Potential energy surface (in meV/f.u) associated with the $B_{oc}$ mode amplitude (in fractional units) starting from a perfectly cubic cell for different doping content x but at fixed octahedra rotation amplitude appearing in the relaxed ground state. An amplitude of 1 corresponds to the amplitude appearing in the ground state of pristine, undoped, SrBiO₃. For each doping content, the cubic cell volume is fixed to the relaxed ground state volume.

***Structural distortions upon K substitutions***: Amplitude of distortions associated with the octahedral rotations $\phi_{xy}^-$ and $\phi_z^+$ as well as the breathing mode $B_{oc}$ at different doping contents are reported in Figure 4.b. The two octahedral rotations amplitude decrease with increasing the K doping content, albeit not disappearing at all in the ground state structure. This fact is due to the A-to-B cation size mismatch that are altered by the introduction of K atoms ($KBiO_3$ have a tolerance factor t=1.01 that favors a cubic cell). The behavior of the breathing mode $B_{oc}$ as a function of the hole doping content is different with respect to rotations: (i) at moderate doping content (x=0 to 0.375), the $B_{oc}$ mode amplitude diminishes until (ii) it totally vanishes at x=0.4375 up to x=0.625. The absence of the breathing mode for 0.375≤x≤0.625 is in line with the similar electronic structures observed for $Bi_S$ and $Bi_L$ cations using the projected density of states presented in Figure 4.a. Thus, Bi cations become all equivalent in the material.

***Doping effects suppress the electronic instability toward disproportionation of the formal oxidation state:*** Potential energy surfaces associated with the breathing mode starting from a perfectly cubic cell for different x values are displayed in Fig. 4.c. Whatever x, the $B_{oc}$ mode is associated with a single well potential (*a>0* in eq 1). It follows that the instability toward disproportionation of the unstable 4+ FOS of cations to more stable 3+/5+ FOS identified in the pristine compound is suppressed by the introduction of K atoms in the material. Furthermore, one can identify that the curvature of the total energy as a function the $B_{oc}$ mode amplitude becomes steeper signaling that the breathing mode hardens with increasing the Sr substitutions. Thus considered alone, the $B_{oc}$ mode possesses an *a* coefficient in eq.1 that becomes positive and larger when x increases.

***Persistence of the breathing mode due to octahedral rotations:*** the coupling of $B_{oc}$ with the two rotations observed in the bulk may still renormalize the effective coefficient $a_{eff}$ in eq.1 by supplying sufficiently large and negative *b* and *c* coefficient contributions counterbalancing the hardening of the $B_{oc}$ mode alone. This is confirmed by the *first-principles* calculations of the potential energy surfaces associated with the $B_{oc}$ mode but at fixed $O_6$ group rotation amplitudes presented in Fig. 4.d. At x=0.25 or x=0.3125, the $B_{oc}$ mode indeed presents a double well potential (*i.e.* $a_{eff}$<0 in eq.1) while at x=0.4375 or x=0.5, it is associated with a single well potential whose minimum is at 0 (*i.e.* $a_{eff}$>0). Thus, up to x=0.375, rotations are sufficiently large to produce a negative effective coefficient $a_{eff}$. At x=0.4375, rotation amplitude are not large enough thereby resulting in a positive effective coefficient $a_{eff}$ in front of $Q_{Boc}^2$ in eq.1.

***Band structure upon K substitutions and the absence of Peierls instability signatures in the superconducting state:*** Figure 5a-g displays the unfolded band structures in the primitive high symmetry *Pm-3m* cubic Brillouin zone of $Sr_{1-x}K_xBiO_3$ for the different doping contents x tested in the simulations. Only bands dispersing around the Fermi level are reported. The pristine material is insulating with an indirect band gap of 0.48 eV. One can notice the existence of gaps along specific path in the Brillouin zone, notably a gap of 1.2 eV halfway the Γ-R path. The existence of this gap comes from the R-point lattice instability producing a dimerization along the [111] cubic direction and the rock-salt pattern of $Bi_L^{3+}$ and $Bi_S^{3+}$ cations in the insulating phase. At low doping content (x=0.0625 and x=0.125), the band structure is mainly not altered but intermediate states appear in the band gap. For doping contents x ranging from 0.1875 to 0.375, the band structure is altered with notably the band gap along the Γ-R path progressively diminishing with increasing x. Although the material is metallic, energy gaps are still present in the band structure and the conduction and valence bands have not yet merged into a single "parabola" centered at the Γ point. This is due to the persistence of the $B_{oc}$ lattice distortion discussed above. At x=0.4375, no more gaps are visible in the band structure and the initial valence and conduction bands in the pristine material have now totally merged. One are then left with a single parabola centered at the Γ point and dispersing on a bandwidth of roughly 4.4 eV. This feature only appears once any finite $B_{oc}$ mode amplitude has totally vanished in the ground state of the material.

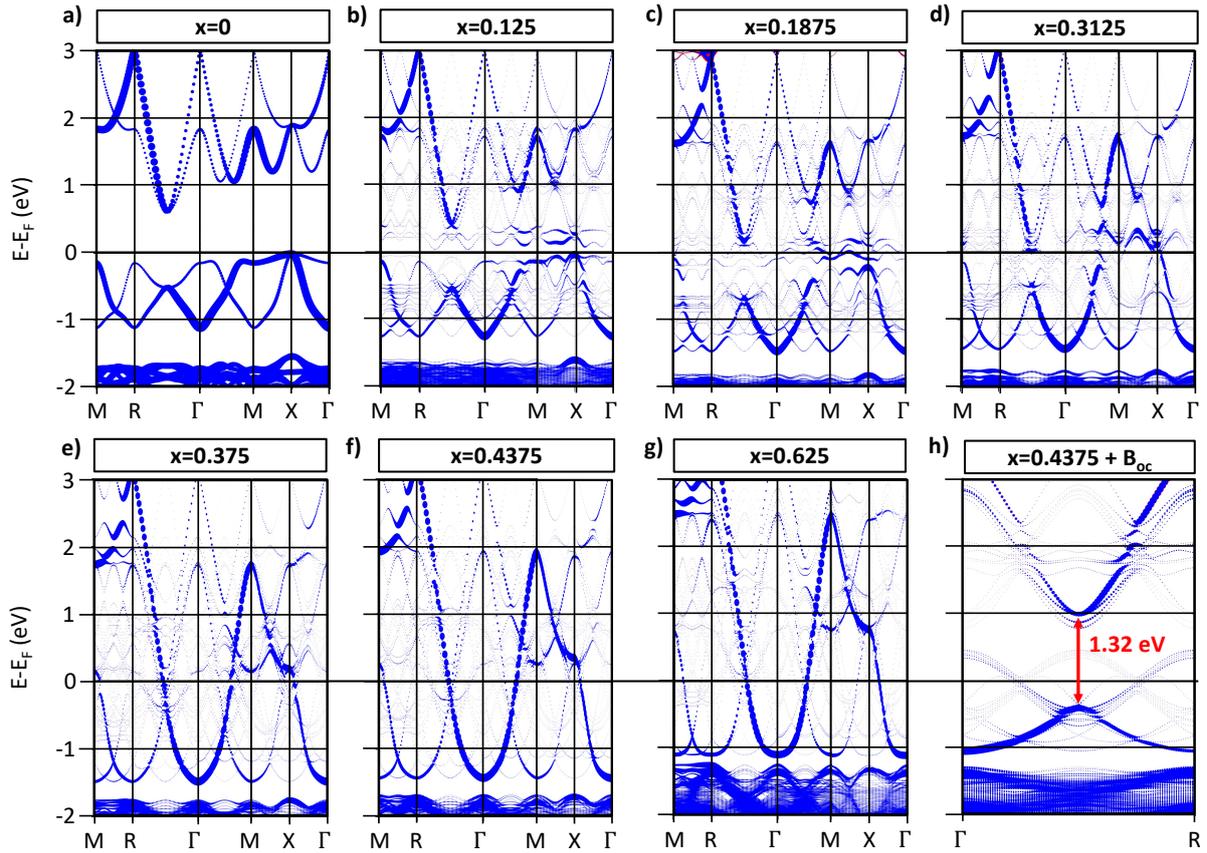

**Figure 5** : **Band structures of doped $Sr_{1-x}K_xBiO_3$. a-g)** Unfolded band structures in the primitive Pm-3m Brillouin zone exhibited by $Sr_{1-x}K_xBiO_3$ as a function of the K doping content x. The size of the dots are associated with the spectral weight of each points. Tiny points are mostly associated with replica associated with the band folding when supercells have to be used. High symmetry reciprocal points have the following coordinates: Γ (0,0,0), X (1/2,0,0), M (1/2,1/2,0) and R (1/2,1/2,1/2). **h)** Gap opening due to the presence of the breathing mode $B_{oc}$ in doped $Sr_{1-x}K_xBiO_3$ at x=0.4375 along the Γ-R path. The breathing mode amplitude is 0.159 Å and the relaxed ground state at x=0.4375 is considered.

**The vicinity of a phase possessing a disproportionation instability is a prerequisite to superconductivity**: the extinction of any type of disproportionation signatures ($Bi_L$-O and $Bi_S$-O bond lengths and $Bi_S$ and $Bi_L$ electronic structure asymmetries) as well as the observation of a single parabola centered at Γ and dispersing over a large energy range for doping contents x ranging from x=0.4375 to x=0.625 in the simulations relates closely with the experimental doping content (x=0.45-0.6) required to reach the superconducting state[10] – one recalls here that the $B_{oc}$ mode is underestimated by ~10% in the DFT bulk with respect to experiments thereby one may predict a superconducting transition at lower doping content with DFT simulations. It clearly suggests that the breathing mode $B_{oc}$ is the key behind the superconducting phase. The latter is likely reached when the effective coefficient in front $Q_{Boc}^2$ in eq.1 is nearly zero or slightly positive. Lattice vibrations with the $B_{oc}$ character potentially facilitated by $O_6$ group rotation vibrations may produce spin paired electrons and holes on the lattice, *i.e.* Cooper pairs, – although the $B_{oc}$ distortion does not stabilize any finite amplitude in the material. One indeed checks in the simulations that the $B_{oc}$ mode at x=0.4375 is strongly softened when rotation amplitudes are fixed to the pristine value ($\emptyset = 1$ in Fig 3.d) with respect to the situation for the ground state at x=0.4375 ($\emptyset = 0.66$ in Fig 3.d). At larger doping content such as x=0.625, vibrations associated with rotations may not sufficiently soften the breathing mode $B_{oc}$. It does explain that although the band dispersion at x=0.625 is similar to the case at x=0.4375, the material is not found superconducting anymore due to a rather hard vibration associated with the breathing mode vibration hindering Cooper pairs formation.

***K doping is mainly related to simple steric effects***: Following the proposed mechanism, the role of the K doping for the appearance of superconductivity is rather indirect and related to a simple steric effect: (i) Sr substitutions by K thus increases the t factor of $Sr_{1-x}K_xBiO_3$ thereby diminishing the rotations amplitude and (ii) in turn it suppresses the breathing mode stabilization in the material.

***The proposed scenario is corroborated by experimental facts in BaBiO₃:*** Ba is larger than Sr atoms and hence the *t* factor of $BaBiO_3$ is larger (t=0.92) than the t factor of $SrBiO_3$. Octahedral rotations in $BaBiO_3$ are ultimately smaller than in $SrBiO_3$ and thus the expected K doping needed to suppress the stabilization of the $B_{oc}$ distortion in the ground state is necessarily smaller in $BaBiO_3$ than in $SrBiO_3$ following the identified mechanism. This is verified experimentally[11]: the superconducting phase in $Ba_{1-x}Bi_xO_3$ is reached for x≈0.30-0.45 while it is for x=0.45-0.6 in $SrBiO_3$.

### 4. Superconducting properties associated with the breathing distortion

***The $B_{oc}$ mode frequency in the superconducting state:*** In order to assess the frequency of the breathing mode $B_{oc}$ in the superconducting phase, the potential energy surface associated with $B_{oc}$ mode is computed by freezing some displacements of the distortion in the

32 f.u. ground state structure (*i.e.* the relaxed DFT cell containing the octahedral rotations) for a doping content x of 0.4375. By fitting the resulting energy curve (see method), the frequency $\omega$ for the breathing mode is computed to 66 meV. This value is in sharp agreement with the quantity extracted experimentally in the superconducting phase of $BaBiO_3$ for this mode ($\omega$=62 meV in Ref.[44]).

**The reduced electron-phonon coupling matrix element:** In order to extract the quantification of the REPME labelled D (see method), a $B_{oc}$ displacement of 0.159 Å is frozen in the ground state structure at x=0.4375. It results in a gap opening of $\Delta E_g$=1.32 eV along the Γ-R path (Figure 5.h). It yields a reduced electron-phonon matrix element (REPME) D=12.4 eV/Å (see method). This value is an improvement over classical LDA and GGA quantities (GGA yields D=7.8 eV/Å) and hybrid DFT using the HSE06 functional (D=11 eV/Å) and closely matches the REPME computed with quasi particle Green's function and screened Coulomb interaction (GW) technic that evaluates D to 13.7 eV/Å (Ref.[3])

**The electron-phonon coupling constant and the critical temperature:** Using this REPME, the computed frequency of the breathing mode and the density of states $N(E_F)$=0.26 states per eV per f.u per spin channel – in sharp agreement with $N(E_F)$ evaluated experimentally to 0.225-0.335 in $Ba_{1-x}K_xBiO_3$, see Ref.[11]—, the electron-phonon coupling (EPC) constant $\lambda$ associated with the breathing mode is evaluated to 1.22. This value is in excellent agreement with the experimental value of $\lambda$=1.3$\pm$ 0.2 obtained in $Ba_{0.51}K_{0.49}BiO_3$ [4] and is an improvement over hybrid DFT calculation of Ref.[3]. The discrepancy with more sophisticated DFT functionals may originate from the fact that important structural distortions and relevant lattice mode couplings were not considered in the ground state calculations of Ref.[3]. The importance of octahedral rotations on the EPC was already suggested in Ref.[28].

Following the methodology proposed in Ref.[3], the logarithmic average frequency is identified to $\omega_{log}$=427K from our SCAN-DFT simulations. Using the Mac Millan equation[45], the critical temperature is estimated between 32 and 25 K for screened Coulomb potentials $\mu^*$of 0.1 and 0.15, respectively (see method). This is in very good agreement with the critical temperatures evaluated in potassium doped bismuthates reaching at most 34K[11].

**DISCUSSION**

The DFT simulations reveal that hole doping has an indirect effect progressively suppressing the charge and bond disproportionation effects through the reduction of the octahedral rotations induced by simple steric effects. Once the disproportionation effects have vanished, no more energy gaps are identified in the band structure and the superconducting phase is reached. In this regime, the bond disproportionation vibration, that can be favored by its coupling with octahedral rotation vibrations, can produce spin paired electrons and holes on the lattice, *i.e.* Copper pairs. The computed electron-phonon and related quantities are all in sharp agreement with experimental quantities obtained in these bismuthates. It is thus clear that the proximity of a charge and bond ordered phase is a prerequisite to superconductivity. These results being ratified with parameter free DFT calculations, it further shows that DFT is sufficient to capture the physics of complex oxide superconductors, thereby showing that strong correlations effects may not be an universal explanation behind high critical temperature observed in oxide superconductors. Finally, bismuthates being very similar to rare-nickelates, it is therefore really questionable if "*the identified mechanism is also relevant for the nickelate superconductors*". The search for similar disproportionation signatures, trends in doping effects and lattice mode couplings between relevant distortions appears as a key point for understanding pairing mechanisms in other oxide superconductors.

**ACKNOWLEDGEMENTS**

JV acknowledges access granted to HPC resources of Criann through the projects 2020005 and 2007013 and of Cines through the DARI project A0080911453.

**METHOD**

***The choice of the exchange-correlation functional***: DFT simulations have been performed with the meta-GGA SCAN[40] functional that improves the correction of self-interaction errors inherent to practice DFT over the classical LDA and GGA functionals and yields correct trends in lattice distortions and metal-insulator transitions as a function of A and B cations in bulk $ABO_3$ perovskite oxides[38]. This functional also has the advantage of being parameter free and can therefore perfectly adapt to multiple formal oxidation states that an ion can develop in a doped material.

***The method to describe doping effects***: Hole doping effects are modelled by substituting $Sr^{2+}$ cations by the monovalent $K^+$ cation. Special Quasi-random Structure (SQS) technique developed by Zunger *et al*[46] that allows to extract the cation arrangement maximizing the disorder characteristic of an alloy within a given supercell size are used to determine the best approximant of the cation disorder at each doping content. Thus, all possible local motifs for Bi cations in terms of surrounding Sr and K cations are available at each doping content, reminiscent of the situation in the real alloy. In order to allow enough flexibility for the material to develop the different lattice distortions exhibited by perovskites that can open the band gap[37], such as those displayed in Figure 1, and to stabilize polaronic states, a 32 formula unit supercell that correspond to a ($2\sqrt{2}$, $2\sqrt{2}$, 4) supercell with respect to the primitive, undistorted, cubic *Pm-3m* cell is used.

***The used crystallographic cell***: Only the low temperature *P2$_1$/n* phase adopted by $SrBiO_3$ is used throughout the study as a starting point.

***Potential energy surfaces and relevant phonon frequency calculation***: Potential energy surfaces associated with lattice distortions are plotted starting from a perfectly undistorted *Pm-3m* cubic cell in which finite amplitudes of the different lattice distortions are condensed. In order to determine the frequencies $\omega$ of interesting lattice distortions in the ground state structure reached upon doping, a full phonon calculation is not affordable for a 32 formula unit supercell without any symmetry. Instead, some lattice mode amplitudes $Q$ associated with a lattice distortion are frozen in the ground state structure and the associated potential energy surface is computed. Recalling that $E = 1/2M\omega^2Q^2$ for an harmonic oscillator with a frequency $\omega$ and where M is the mass of the moving atoms, the frequency is identified by fitting the potential energy surface with an expression, of the $E = aQ^2 + bQ^4$ where a and b are coefficients representing harmonic and anharmonic effects. One can then identify that $\omega = \sqrt{\frac{2a}{M}}$.

***Computing superconducting properties***: the reduced electron-phonon coupling matrix element (REPME) associated with the breathing distortion $B_{oc}$ are evaluated by freezing its atomic displacements in the relaxed ground state structure. Using the gap opening amplitude $\Delta E_g$ appearing in the band structure due to the frozen phonon displacement, the REPME labelled D is computed by the following formula :

$$D = \frac{\Delta E_g}{2u} \quad (eq.\ 2)$$

where u is the displacement of one oxygen atom. The electron-phonon coupling constant $\lambda$ is evaluated by the following equation:

$$\lambda = N(E_F)\left(\frac{\hbar^2}{2M\omega_{Boc}^2}\right)D^2 \quad \text{(eq. 3)}$$

where $N(E_F)$ is the density of states at the Fermi level per spin channel per formula unit, M is the mass of the displaced atoms and $\omega_{Boc}^2$ is the square of the computed $B_{oc}$ frequency. The critical temperature $T_c$ is computed by using the modified Mac Millan equation[45]:

$$T_c = \frac{\omega_{log}}{1.2}\exp\left(-\frac{1.04(1+\lambda)}{\lambda - \mu^*(1+0.62\lambda)}\right)$$

where $\omega_{log}$ is average logarithmic frequency and $\mu^*$ is the screened Coulomb potential with conventional values between 0.1 and 0.15.

***Other technical details***: DFT simulations are performed with the Vienna Ab initio Simulation package (VASP)[47,48]. The energy cut-off is set to 650 eV and is accompanied by a 4x4x3 Gamma-centred kmesh for the 32 f.u. supercell. The kmesh is increased to 5x5x3 for density of states, potential energy surfaces and frequency calculations. Projector Augmented Wave (PAW) potentials[49] are used with Bi *d* states being treated as core states. Geometry relaxations (atomic positions plus lattice parameters) are performed until forces acting on each atom are lower than 0.05 eV/Å. Symmetry of the relaxed structures are extracted with the Findsym application[50] and amplitudes of distortions are extracted using a symmetry mode analysis with respect to the primitive *Pm-3m* cell with the Isodistort tool from the Isotropy applications[51,52].

**References**

1. Cooper, L. N. Bound Electron Pairs in a Degenerate Fermi Gas. *Phys. Rev.* **104**, 1189 (1956).
2. Li, D. *et al.* Superconductivity in an infinite-layer nickelate. *Nature* **572**, 624–627 (2019).
3. Yin, Z. P., Kutepov, A. & Kotliar, G. Correlation-enhanced electron-phonon coupling: Applications of GW and screened hybrid functional to bismuthates, chloronitrides, and other high-Tc superconductors. *Phys. Rev. X* **3**, 021011 (2013).
4. Wen, C. H. P. *et al.* Unveiling the Superconducting Mechanism of Ba0.51 K0.49BiO3. *Phys. Rev. Lett.* **121**, 117002 (2018).

5.  Keimer, B., Kivelson, S. A., Norman, M. R., Uchida, S. & Zaanen, J. From quantum matter to high-temperature superconductivity in copper oxides. *Nature* **518**, 179–186 (2015).

6.  Garg, A., Randeria, M. & Trivedi, N. Strong correlations make high-temperature superconductors robust against disorder. *Nat. Phys.* **4**, 762 (2008).

7.  Li, H. *et al.* Coherent organization of electronic correlations as a mechanism to enhance and stabilize high-T C cuprate superconductivity. *Nat. Commun.* **9**, 26 (2018).

8.  Worm, P. *et al.* Correlations turn electronic structure of finite-layer nickelates upside down. *ArXiv* 211.122697v2 (2021).

9.  Kitatani, M. *et al.* Nickelate superconductors—a renaissance of the one-band Hubbard model. *npj Quantum Mater.* **5**, 59 (2020).

10. Chaillout, C. *et al.* Discovery of a second family of bismuth-oxide-based superconductors. *Nature* **390**, 148 (1997).

11. Sleight, A. W. Bismuthates: BaBiO3 and related superconducting phases. *Phys. C Supercond. its Appl.* **514**, 152–165 (2015).

12. Mattheiss, L. F., Gyorgy, E. M. & Johnson, D. W. Superconductivity above 20 K in the Ba-K-Bi-O system. *Phys. Rev. B* **37**, 3745 (1988).

13. Federici, J. F., Greene, B. I., Hartford, E. H. & Hellman, E. S. Optical characterization of excited states in BaBiO3. *Phys. Rev. B* **42**, 923 (1990).

14. Sarkar, S., Raghunathan, R., Chowdhury, S., Choudhary, R. J. & Phase, D. M. The Mystery behind Dynamic Charge Disproportionation in BaBiO3. *Nano Lett.* **21**, 8433 (2021).

15. Sleight, A. W., Gillson, J. L. & Bierstedt, P. E. High-temperature superconductivity in the BaPb1-xBixO3 system. *Solid State Commun.* **17**, 27 (1975).

16. Kennedy, B. J., Howard, C. J., Knight, K. S., Zhang, Z. & Zhou, Q. Structures and phase transitions in the ordered double perovskites Ba2BiIIIBiVO6 and Ba2BiIIISbVO6. *Acta Crystallogr. B.* **62**, 537 (2006).

17. Glazer, A. M. The classification of tilted octahedra in perovskites. *Acta Crystallogr. Sect. B Struct. Crystallogr. Cryst. Chem.* **28**, 3384–3392 (1972).

18. Goldschmidt, V. M. Die Gesetze der Krystallochemie. *Naturwissenschaften* **14**, 477–485 (1926).

19. Dalpian, G. M., Liu, Q., Varignon, J., Bibes, M. & Zunger, A. Bond disproportionation,

charge self-regulation, and ligand holes in s- p and in d-electron ABX3 perovskites by density functional theory. *Phys. Rev. B* **98**, 075135 (2018).

20. Thonhauser, T. & Rabe, K. M. Fcc breathing instability in BaBiO3 from first principles. *Phys. Rev. B* **73**, 212106 (2006).

21. Meregalli, V. & Savrasov, S. Y. Electron-phonon coupling and properties of doped BaBiO3. *J. Supercond. Nov. Magn.* **12**, 185 (1999).

22. Foyevtsova, K., Khazraie, A., Elfimov, I. & Sawatzky, G. A. Hybridization effects and bond disproportionation in the bismuth perovskites. *Phys. Rev. B* **91**, 121114(R) (2015).

23. Varignon, J., Santamaria, J. & Bibes, M. Electrically Switchable and Tunable Rashba-Type Spin Splitting in Covalent Perovskite Oxides. *Phys. Rev. Lett.* **122**, 116401 (2019).

24. Zaanen, J., Sawatzky, A. & Allen, J. W. Band Gaps and Electronic Structure of Transition-Metal Compounds. *Phys. Rev. Lett.* **55**, 418 (1985).

25. Varignon, J., Malyi, O. I. & Zunger, A. Dependence of band gaps in d -electron perovskite oxides on magnetism. *ArXiv* 1–14

26. Khazraie, A., Foyevtsova, K., Elfimov, I. & Sawatzky, G. A. Oxygen holes and hybridization in the bismuthates. *Phys. Rev. B* **97**, 075103 (2018).

27. Franchini, C., Sanna, A., Marsman, M. & Kresse, G. Structural , vibrational , and quasiparticle properties of the Peierls semiconductor BaBiO 3 : A hybrid functional and self-consistent GW + vertex-corrections study. *Phys Rev B* **81**, 085213 (2010).

28. Bazhirov, T., Coh, S., Louie, S. G. & Cohen, M. L. Importance of oxygen octahedra tilts for the electron-phonon coupling in K-doped BaBiO 3. *Phys. Rev. B* **88**, 224509 (2013).

29. Franchini, C., Kresse, G. & Podloucky, R. Polaronic Hole Trapping in Doped BaBiO 3. *Phys Rev Lett.* **102**, 256402 (2009).

30. Benam, M. R., Foyevtsova, K., Khazraie, A., Elfimov, I. & Sawatzky, G. A. Bond versus charge disproportionation and nature of the holes in s – p A B X 3 perovskites. *Phys. Rev. B* **104**, 195141 (2021).

31. Smolyanyuk, A., Boeri, L. & Franchini, C. Ab initio prediction of the high-pressure phase diagram of BaBiO3. *Phys. Rev. B* **96**, 035103 (2017).

32. Rice, T. M. & L., S. Real-Space and k-Space Electron Pairing in BaPb1-xBix03. **47**, 689 (1981).

33. Varma, C. M. Missing valence states, diamagnetic insulators, and superconductors.

*Phys. Rev. Lett.* **61**, 2713 (1988).

34. Chakraverty, B. K. & Ranninger, J. Bipolarons and superconductivity. *Philos. Mag. Part B* **52**, 669 (1985).

35. Scalapino, D. J., White, S. R. & Zhang, S. Insulator, metal, or superconductor: The criteria. *Phys. Rev. B* **47**, 7995 (1993).

36. Zeyher, R. & Kunc, K. Instabilities in cubic BaBiO3 from total-energy calculations. *Solid State Commun.* **74**, 805 (1990).

37. Varignon, J., Bibes, M. & Zunger, A. Origin of band gaps in 3d perovskite oxides. *Nat. Commun.* **10**, 1658 (2019).

38. Varignon, J., Bibes, M. & Zunger, A. Mott gapping in 3dABO3 perovskites without Mott-Hubbard interelectronic repulsion energy U. *Phys. Rev. B* **100**, 035119 (2019).

39. Varignon, J., Bibes, M. & Zunger, A. Origins Vs. fingerprints of the Jahn-Teller effect in d-electron ABX3 perovskites. *Phys. Rev. Res.* **1**, 033131 (2019).

40. Sun, J., Ruzsinszky, A. & Perdew, J. Strongly Constrained and Appropriately Normed Semilocal Density Functional. *Phys. Rev. Lett.* **115**, 036402 (2015).

41. Iglesias, L., Bibes, M. & Varignon, J. First-principles study of electron and hole doping effects in perovskite nickelates. *Phys. Rev. B* **104**, 035123 (2021).

42. Furness, J. W. *et al.* An accurate first-principles treatment of doping-dependent electronic structure of high-temperature cuprate superconductors. *Commun. Phys.* **1**, 11 (2018).

43. Mercy, A., Bieder, J., Íñiguez, J. & Ghosez, P. Structurally triggered metal-insulator transition in rare-earth nickelates. *Nat. Commun.* **8**, 1677 (2017).

44. Braden, M., Reichardt, W., Ivanov, A. S. & Rumiantsev, A. Y. Anomalous dispersion of LO phonon branches in Ba0.6K0.4BiO3. *Europhys. Lett.* **34**, 531 (1996).

45. Allen, P. B. & R. C. Dynes. Transition temperature of d-f-band superconductors. *Phys. Rev. B* **12**, 905 (1975).

46. Zunger, A., Wei, S.-H., Ferreira, L. G. & Bernard, J. E. Special quasirandom structures. *Phys. Rev. Lett.* **65**, 353 (1990).

47. Kresse, G. & Haffner, J. No Title. *Phys. Rev. B* **47**, 558 (1993).

48. Kresse, G. & Furthmüller, J. Efficiency of ab-initio total energy calculations for metals and semiconductors using a plane-wave basis set. *Comput. Mater. Sci.* **6**, 15 (1996).

49. Blöchl, P. E. Projector augmented-wave method. *Phys. Rev. B* **50**, 17953 (1994).

50. Stokes, H. T. & Hatch, D. M. FINDSYM: Program for identifying the space-group symmetry of a crystal. *J. Appl. Crystallogr.* **38**, 237 (2005).

51. Campbell, B. J., Stokes, H. T., Tanner, D. E. & Hatch, D. M. ISODISPLACE: A web-based tool for exploring structural distortions. *J. Appl. Crystallogr.* **39**, 607–614 (2006).

52. ISOTROPY Software Suite, iso.byu.edu. Available at: https://iso.byu.edu/iso/isodistort_version5.6.1/isodistort.php.